\documentclass[journal]{IEEEtran}
\pdfoutput=1
\usepackage[nolist]{acronym}
\usepackage{subfig}
\usepackage{graphicx}
\usepackage[usenames]{color}
\usepackage[dvipsnames]{xcolor}
\usepackage{amsmath}
\usepackage{uppaal}
\usepackage{bm}
\usepackage{multicol}
\usepackage{enumitem}

\PassOptionsToPackage{hyphens}{url}
\usepackage{url}

\newcommand{\cmdtt}[1]{\texttt{\textbf{\detokenize{#1}}}}

\usepackage[acronym]{glossaries}
\newacronym{TG}{TG}{Task Group}
\newacronym{AVB}{AVB}{Audio Video Bridging}
\newacronym{TSN}{TSN}{Time-Sensitive Networking}
\newacronym{CDES}{CDES}{critical distributed embedded systems}
\newacronym{SRP}{SRP}{Stream Reservation Protocol}
\newacronym{CSRP}{CSRP}{Consistent Stream Reservation Protocol}
\newacronym{TA}{TA}{Talker Advertise}
\newacronym{TF}{TF}{Talker Failed}
\newacronym{LAF}{LAF}{Listener Asking Failed}
\newacronym{LR}{LR}{Listener Ready}
\newacronym{LRF}{LRF}{Listener Ready Failed}
\newacronym{FIFO}{FIFO}{first-in, first-out}
\newacronym{FD}{FD}{Final Decision}
\newacronym{LNR}{LNR}{List of Nodes Resources}
\newacronym{QoS}{QoS}{quality of service}

\begin{document}

\title{Description of the \uppaal Models for SRP and CSRP and Verification of their Termination and Consistency Properties}

\author{\IEEEauthorblockN{Daniel Bujosa\IEEEauthorrefmark{1}, In\'{e}s \'{A}lvarez\IEEEauthorrefmark{2}, Juli\'{a}n Proenza\IEEEauthorrefmark{2}}\\
\IEEEauthorblockA{\IEEEauthorrefmark{1}M{\"{a}}lardalen University, V{\"{a}}ster{\aa}s, Sweden\\
\IEEEauthorrefmark{2} University of the Balearic Islands, Palma, Spain}\\
daniel.bujosa.mateu@mdh.se, \{ines.alvarez, julian.proenza\}@uib.es}

\maketitle

\begin{abstract}
    The IEEE Audio Video Bridging (AVB) Task Group (TG) was created to provide Ethernet with soft real-time guarantees. Later on, the TG was renamed to Time-Sensitive Networking (TSN) and its scope broadened to support hard real-time and critical applications. The Stream Reservation Protocol (SRP) is a key work of the TGs as it allows reserving resources in the network, guaranteeing the required quality of service (QoS). AVB's SRP is based on a distributed architecture, while TSN's is based on centralized ones. The distributed version of SRP is supported and used in TSN. Nevertheless, it was not designed to provide properties that are important for critical applications. Therefore, we propose a new version of \gls{SRP} with enhanced services called \gls{CSRP}. In this document we describe the \gls{SRP} and \gls{CSRP} \uppaal models we developed and the queries we used to verify their termination and consistency properties. 
    
\end{abstract}

\section{Introduction}\label{Intro}

In this document we describe the \gls{SRP} and \gls{CSRP} \uppaal models we developed and the queries we used to verify their termination and consistency properties. In Section \ref{SRP_Overview} we describe the operation of \gls{SRP}. In Section \ref{UPPAAL_Overview} we explain the most relevant characteristics of \uppaal.
 In Sections \ref{U_SRP_M}, \ref{TE} and \ref{CE} we describe the \gls{SRP} model and the verification of the termination and consistency properties respectively. In Section \ref{SP} we describe the operation of \gls{CSRP} and in Sections \ref{CSRP_model}, \ref{CSRP_TE} and \ref{CSRP_CE} we describe the \gls{CSRP} model and the verification of the termination and consistency properties respectively.

\section{\gls{SRP} Overview}\label{SRP_Overview}

\gls{SRP} follows the publisher-subscriber paradigm, where the publisher is called talker and the subscribers, listeners. The real-time data communications are made through streams. A stream is a logical communication channel that carries traffic defined by a set of parameters, such as the period or frame size.



When a talker wants to transmit a set of frames with certain parameters, it must first create the stream to convey such frames. To create a stream the talker has to declare its intention to communicate by transmitting in broadcast mode a special message called \gls{TA} message. This message conveys stream identification information, as well as the resources needed to convey the traffic. This information is then used in the rest of devices of the network to know whether there are enough resources for the stream so that it can be created. Notice that \gls{SRP} relies on other mechanisms that eliminate the loops in the network to prevent the \gls{TA} message from circulating the network indefinitely.

The \gls{TA} message transmitted by the talker is received by the bridge connected to it. When a bridge receives a \gls{TA} message, every forwarding port (all ports through which the \gls{TA} message was not received) checks if it has enough resources for the stream or not. If the port has enough resources, the TA message is forwarded to the next device, i.e., the next bridge or a node. On the other hand, if the port does not have enough resources, it sends a so called \gls{TF} message instead. A \gls{TF} message conveys the same information as the \gls{TA} message plus the reason for the failure in the reservation. Bridges that receive a \gls{TA} message transmitted by another bridge through one of their ports behave as we have just described. On the contrary, if the message received is a \gls{TF} message, bridges transmit a \gls{TF} message through all their forwarding ports.

Regarding nodes, we have to note that not all nodes are listeners for all streams. Therefore, if a node that does not want to become a listener of the stream receives a \gls{TA} or \gls{TF} message, it does not carry any further actions. In fact, it does not even inform the talker about its lack of interest in the stream. On the other hand, if a node receives a \gls{TA} or \gls{TF} message and is willing to listen to the stream there are three possible scenarios to consider: (i) the listener receives a \gls{TF} message and cannot therefore receive, so it sends a message called \gls{LAF} to the bridge; (ii) the listener receives a \gls{TA} message but, while checking its resources it realises that it does not have enough resources to receive the stream, so it sends an \gls{LAF} message to the bridge; and, (iii) the listener receives a \gls{TA} message and, while checking its resources it realises that it has enough resources to receive the stream, so it sends a message called \gls{LR} message to the bridge.

The port of the bridge connected to the listener can receive an \gls{LR} or \gls{LAF} message. If the port receives an \gls{LAF} message it does nothing else. If the port receives an \gls{LR} message the port checks its resources again. If it does not have enough resources the port changes the \gls{LR} received to an \gls{LAF}; otherwise, if it has enough resources, the port reserves the resources. Whenever a bridge has several listener responses to forward, it combines the responses into a single one and transmits it to the talker. The result of combining the responses is the following: (i) if the bridge receives an \gls{LR} in all the ports, it transmits to the talker an \gls{LR} message; if the bridge receives an \gls{LAF} in all the ports, it transmits to the talker another \gls{LAF} message; and, if the bridge receives \gls{LR} messages in some ports and \gls{LAF} messages in other ports, it will transmits to the talker a new message called \gls{LRF} message. Whenever a bridge receives an \gls{LRF} message it forwards an \gls{LRF} message to the talker, regardless of the other listener attributes it receives.

Finally, waits until it receives an \gls{LR} or \gls{LRF} message to start the data transmission. Once the stream has been created, the talker can delete it at any time by means of the unadvertise stream mechanism. The talker transmits a message to eliminate the stream from all devices. This message is also transmitted in broadcast mode to ensure that all bridges and listeners receive the indication to eliminate the stream.


\section{\uppaal Overview}\label{UPPAAL_Overview}

The \uppaal model checker is a tool for modelling real-time systems and formally verify their properties \cite{Behrmann2004}.
In \uppaal the systems are modelled by means of interconnected timed automata (finite-state machines extended with clock that progress at the same pace). In addition, \uppaal provides a formal query language that allows defining properties that the system should have. Using the model and the queries as inputs, the model performs a exhaustive check of the properties i.e. it explores all the possible execution paths of the model to check whether the properties hold. After this, \uppaal informs the user about the result and, if a property does not hold, it shows an execution path in which the property is violated. We next describe the modelling tool and the query language in more detail.

\subsection{Modelling Tool}

As said before, in \uppaal the systems are modelled as a network of timed automata. Each automaton is specified by a template that can be instantiated several times. At the same time, templates are constructed using locations, edges, local variables and local clocks, and can synchronise through different types of channels. The combination of the activated locations, the value of the variables and the time of the system defines the states which are exhaustively analysed by using queries as we will see in the next sub-section.

Each automaton progresses through a set of locations. There are three different kind of locations: normal, urgent and committed. The difference lies in the time that an automaton can remain in it. An automaton can remain indefinitely in a normal location, unless the residence time is limited using invariants. On the other hand, an automaton must immediately leave any urgent or committed location, that is, the time that an automaton can remain in such a location is 0 and, therefore, the time does not pass in this type of locations. In this sense, the difference between committed and urgent is that the committed locations are atomic, while the urgent locations are not. Atomic means that they are not affected by the actions carried out by other automata with locations of the same type. On the other hand, a location of each automata should also be an initial location, which is the location where the automaton will start operating.

As we have said, the time an automaton can remain in a normal location can be bounded by means of invariants. An invariant is an expression placed in the locations that impose a condition to remain in it. For example, if one location has the invariant \cmdtt{t <= 5}, the automaton has to leave the location as soon as the variable t, which can be a clock or any other kind of variable, becomes greater than 5.

An automaton can move through their locations using the edges. The edges can be enable or disabled by using guards. The guards are expressions defined by variables and clocks which disable the automaton to take the corresponding edge, if the expression is not true. In addition, edges can assign values to the variables when they are taken.

Finally, automata can synchronise each other by taking certain edges at the same time. This can be done by using the already mentioned channels. The channels are variables which can be labelled in the edges. There are different kind of channels but in this work we only used two of them so these will be the ones we explain. The first type of channel we will explain is the binary channel. In this kind of channels there are only two edges labelled for each channel variable. These two edges can only be taken at a time, so they will wait for each other to be taken. The other type of channel we will explain is the broadcast channel. In this kind of channels one edge is labelled as sender while one or more edges are labelled as receivers. Receivers have to wait for the sender to be taken while sender edges can be taken at any time, even if not all or none of the receivers are waiting.

\subsection{Query Language}

Queries are expressions used to analyse the model and they are formed by two parts: the state formula and the path formula. The state formulae (represented by \(\bm{\varphi}\)) are expression that can be true or false depending on the state of the model. For example, the state formula \cmdtt{i == 7} is only true in the states in which the value of the variable i is equal to 7. On the other hand, the path formulae are expressions that can be true or false depending on the distribution of the states in which the state formula is met.

\begin{figure}[t]
\centering
\subfloat[Representation of the reachability property.]
{	
	\includegraphics[scale=0.175]{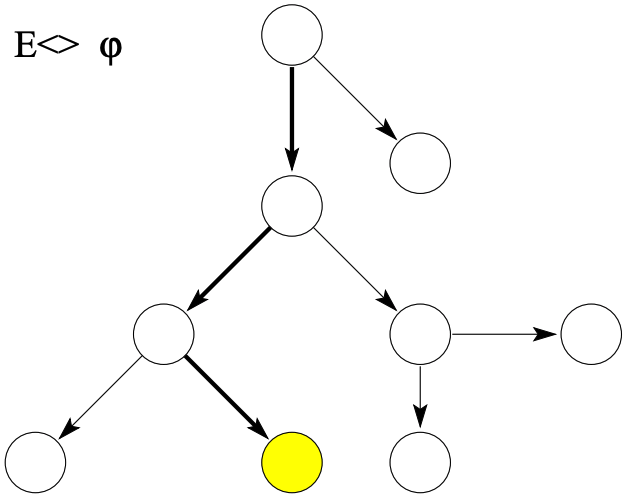}
	\label{fig:e<>}
}\hfill
\subfloat[Representation of the first safety property.]
{	
	\includegraphics[scale=0.175]{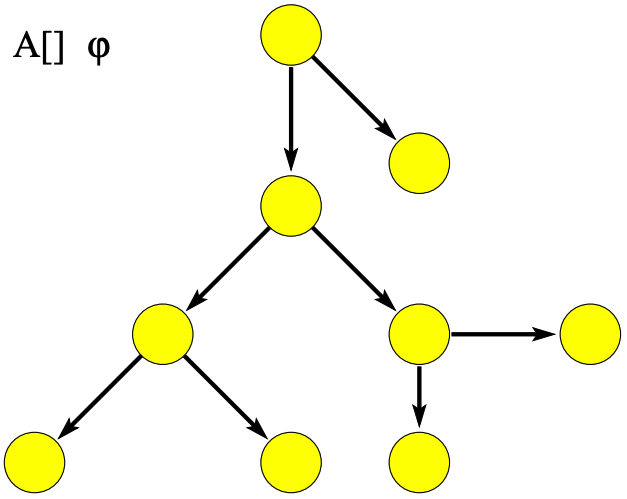}
	\label{fig:a[]}
}

\subfloat[Representation of the second safety property.]
{	
	\includegraphics[scale=0.175]{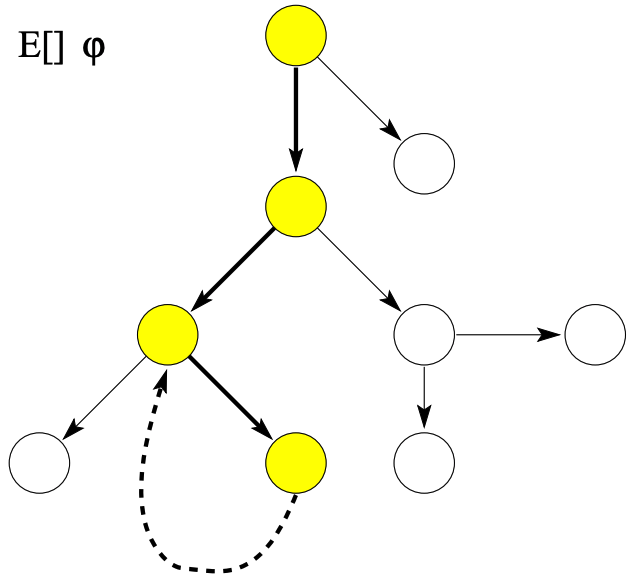}
	\label{fig:e[]}
}\hfill
\subfloat[Representation of the first liveness property.]
{	
	\includegraphics[scale=0.175]{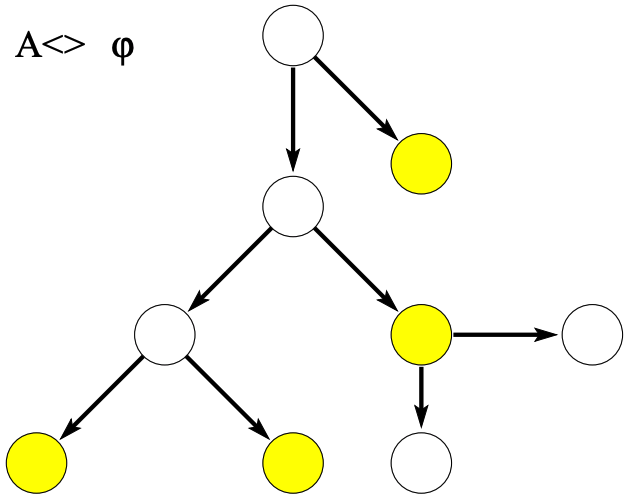}
	\label{fig:a<>}
}\hfill
\subfloat[Representation of the second liveness property.]
{	
	\includegraphics[scale=0.24]{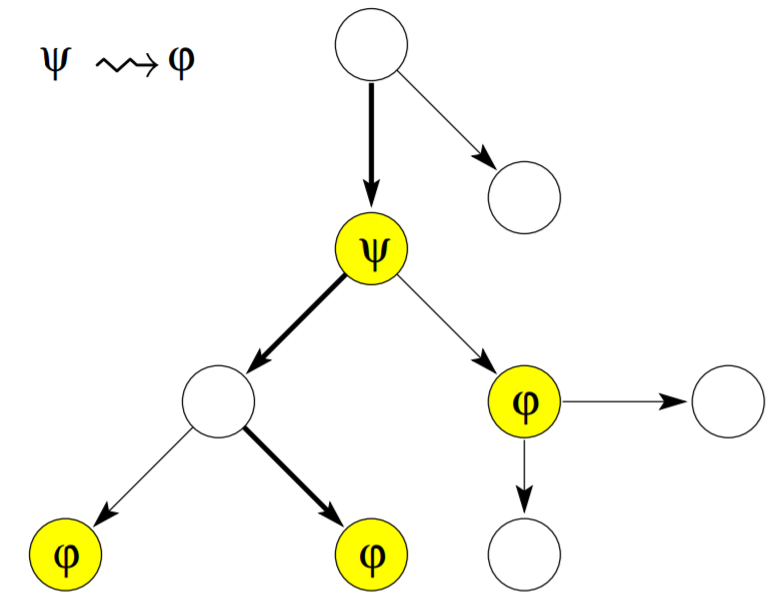}
	\label{fig:a<>2}
}
\caption{Properties that can be evaluated in \uppaal based 
on a figure from~\cite{Behrmann2004}. Each figure shows the paths for which the 
state formula holds; whereas the filled states are the ones where the state 
formulae is satisfied.}\label{fig:uppaal-properties}
\end{figure}

There are 5 types of path formula depending on the already mentioned distribution of the states which can be classified into 3 kind of properties. \figurename\ref{fig:uppaal-properties} shows a graphic representation of these queries. In this figure, circles represent states of the model whereas yellow ones represent states in which the state formula is satisfied (\figurename\ref{fig:a<>2} also ads a symbol into the yellow bubbles to identify which state formula is satisfied in which state). Additionally, the bold arrow shows the path analysed by the path formula. The first property is the reachability property. It checks if exist any state in which the state formula is met and its corresponding state formula is \cmdtt{E<>}\(\bm{\varphi}\). The second one is the safety property. It checks if in all the states (by means of the path formula \cmdtt{A[]}\(\bm{\varphi}\)) or in, at least, a path of the state space (by means of the path formula \cmdtt{E[]}\(\bm{\varphi}\)) the state formula is met. Finally, the third property is the liveness property. It checks if the state formula is eventually met by using the path formula \cmdtt{A<>}\(\bm{\varphi}\) while with the path formula \(\bm{\Psi}\)\cmdtt{-->}\(\bm{\varphi}\) it checks if the state formula \(\bm{\varphi}\) is eventually met after a state in which the state formula \(\bm{\Psi}\) was met.

\section{\gls{SRP} \uppaal Model}\label{U_SRP_M}

\begin{figure}[!t]
\centering
\includegraphics[width=\columnwidth]{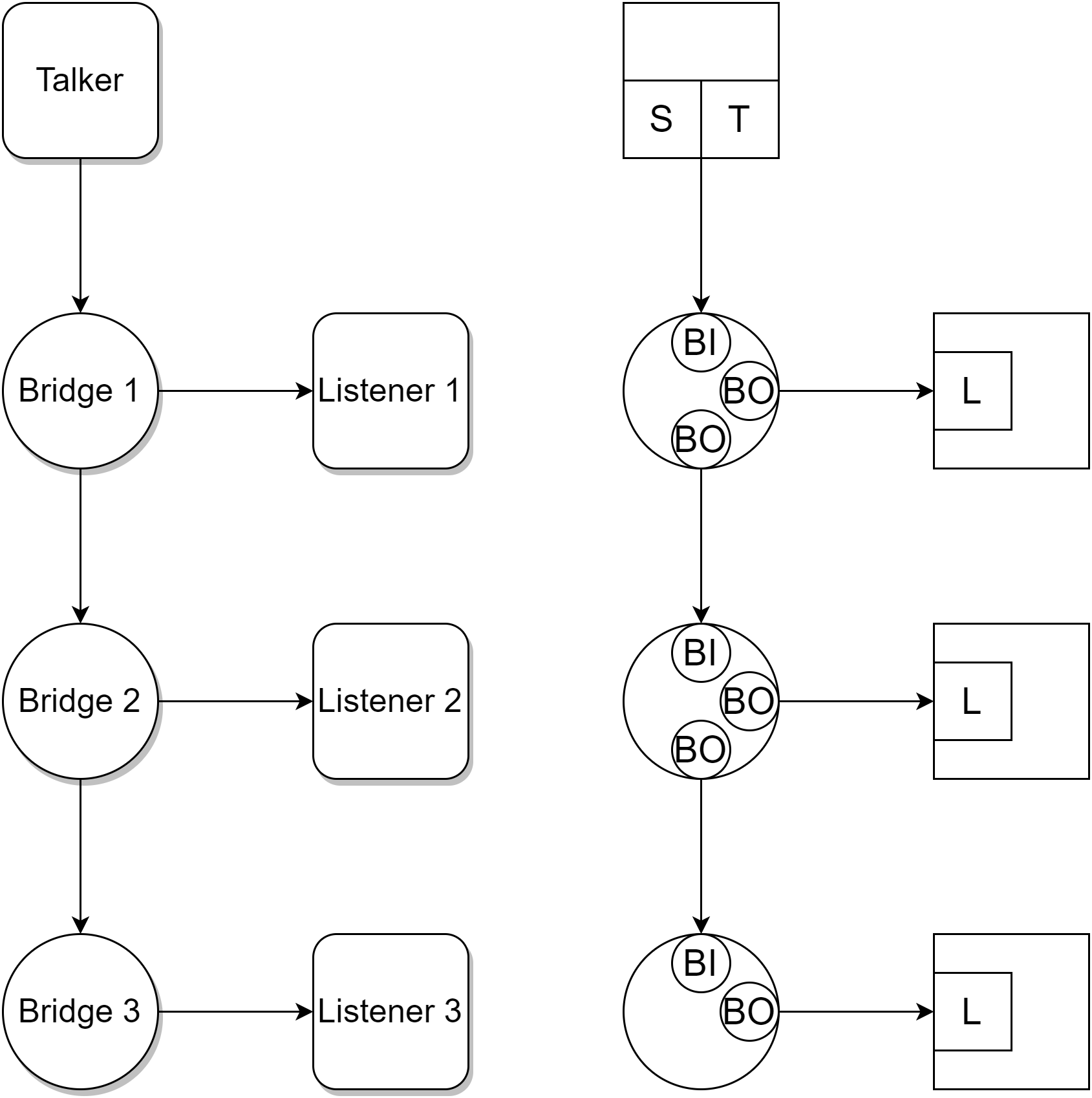}
\begin{multicols}{2}
\begin{enumerate}[label=(\alph*)]
\item Modelled network consisting of one talker, three listeners and three bridges.\columnbreak
\item Abstraction of the network model made with \uppaal.
\end{enumerate}
\end{multicols}
\caption{Representation of the modelled network and its model by means of templates where T represents the Talker template, S the Stream template, BI the BridgeInput template, BO the BridgeOutput template and L the Listener template.}
\label{fig:topology}
\end{figure}

This section introduces the \gls{SRP} model developed in this work.
\figurename \ref{fig:topology} (a) represents the network we modelled with \uppaal while \figurename \ref{fig:topology} (b) represents the resulting \uppaal model. As we can see, our \gls{SRP} model is made of 5 different templates: Talker template, Stream template, Listener template, BridgeInput template and BridgeOutput template (represented as T, S, L, BI, BO respectively in \figurename \ref{fig:topology}(b)). These templates model the different relevant actions of the protocol carried out by the talkers, bridges and listeners. Specifically, as we can see in \figurename \ref{fig:topology}, our model has one instantiation of the Talker and Stream templates to model the actions carried out by one talker. It also has three instantiations of the Listener template to model the actions carried out by three listeners. And, finally, it has 3 instantiations of the BridgeInput template and five instantiations of the BridgeOutput template to model the actions carried out by three \gls{AVB} bridges. Other network elements, such as links, are represented in the model by variables, clocks and channels.

As can be seen in \figurename \ref{fig:topology}(a), and as we have already said, our model is made up of one talker, three bridges and three listeners, each listener connected to one bridge. We decided to use three listeners for many reasons. The first reason is that, in many systems is usual to use active replication, using three replicas which perform majority vote on each result, in order to tolerate the failure of nodes. Moreover, three listeners are enough to have all relevant combinations of responses of the listeners.
On the other hand, we connected one listener to each bridge to have paths with different lengths and end-to-end delays, factors that increase the likelihood of encountering consistency issues. Finally, we used a line topology because \gls{SRP} relies on other protocols to eliminate the loops of the network, such as the Rapid Spanning Tree Protocol \cite{standardd2004} or the Shortest Path Bridging Protocol \cite{standardq2012}.

Like any model of a system, our \gls{SRP} model has a series of abstractions that we describe next. First, we only model the transmission of one stream because allowing the model to transmit several streams would lead to the explosion of the state space without providing any benefit, on the contrary, it would make the model more difficult to analyse. We neither model the transmission of data frames because it is not part of \gls{SRP} and it would increase the complexity of the model unnecessarily, as it would distort the model without giving
greater precision to the analysis of the protocol.
Finally, we did not take into account the presence of errors for several reasons. First, the property issues we detected appear in the absence of faults in the network. Secondly, there are some works like the one presented in \cite{ines2017} that allow tolerating faults in the channel by using proactive replication of frames.

In this work we present a detailed, yet analysable and general model. Specifically, our model divides the Bridge template into two, one for the reception port of the talker attributes and transmission of the listener attributes and another for the reception of the listener attributes and transmission of the talker attributes. These templates can be instantiated as many times as necessary for each bridge, so the generality of the model is maintained. 
Next the templates are described in detail.

\subsection{Talker Templates}

\begin{figure*}[!t]
\centering
\includegraphics[width=\textwidth]{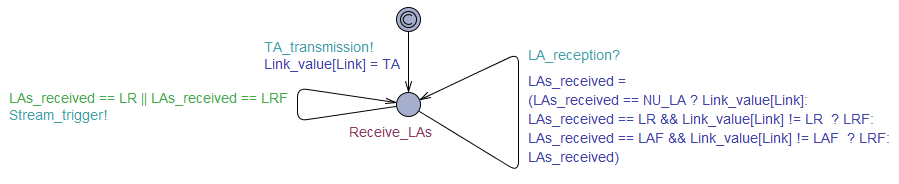}
\caption{Talker template.}
\label{fig:TT}
\end{figure*}

\begin{figure}[!t]
\centering
\includegraphics[width=\columnwidth]{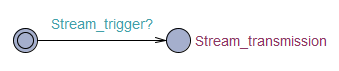}
\caption{Stream template.}
\label{fig:ST}
\end{figure}

The templates of the talker are shown in Figures \ref{fig:TT} and \ref{fig:ST}. The first and most complex is the one that performs the main actions of \gls{SRP} in the talker while the second represents the transmission of data frames.

Since \gls{SRP} begins with the talker's declaration of its intention to transmit, the talker's template begins at a location that, apart from being initial, it is also committed to prevent a deadlock from occurring right at the beginning of the execution. After transmitting the \gls{TA} message, the talker goes to a location where it receives the listeners' responses and, if possible, triggers the stream transmission. The first action is performed in the loop on the right of the automaton, while the second action is performed on the left edge. This last edge does not form a loop since it can only be taken once because the stream must only be triggered once. Taking the edge on the left causes the Stream template to transition to the Stream-transmission location, which represents the stream transmission. After this, the talker can continue receiving listener responses.

\subsection{Listener Template}

\begin{figure}[!t]
\centering
\includegraphics[width=0.5\columnwidth]{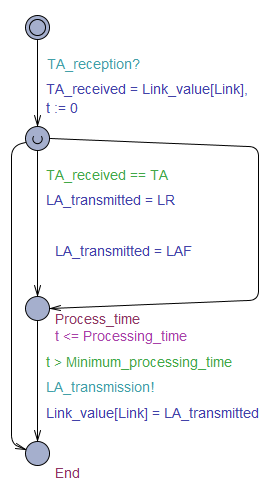}
\caption{Listener template.}
\label{fig:LT}
\end{figure}

The template of the listener is shown in Figure \ref{fig:LT}. It performs the main actions of \gls{SRP} in the listener.

The Listener template starts waiting the reception of a talker attribute (\gls{TA} or \gls{TF}). After receiving it, the listener can perform 3 different actions. If the listener is not interested in the stream the template will take the left edge to the end location and it will not perform any other action. If the listener has received a \gls{TA} message and has enough resources, it will take the central edge and will prepare an \gls{LR} message to be transmitted. Otherwise, it will take the right edge and will prepare an \gls{LAF} message to be transmitted. The template will remain in the Process\_time location an undefined time between 10 and 200 ms. The first value is the minimum process time we measured in a real experimental setup, while the second value is the maximum time that a message can take to be transmitted according to the \gls{AVB} standard \cite{standardq2012}. After this processing time, the template will take the edge to the End location while transmitting listener response (\gls{LR} or \gls{LAF} message).

\subsection{Bridge Templates}

\begin{figure*}[!t]
\centering
\includegraphics[width=0.6\textwidth]{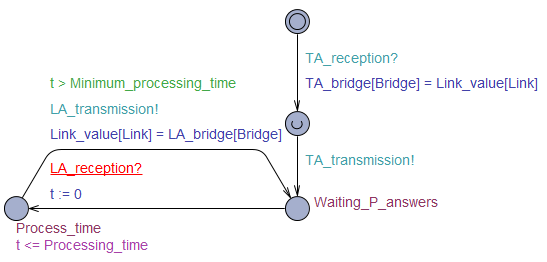}
\caption{BridgeInput template.}
\label{fig:BIT}
\end{figure*}

\begin{figure*}[!t]
\centering
\includegraphics[width=0.9\textwidth]{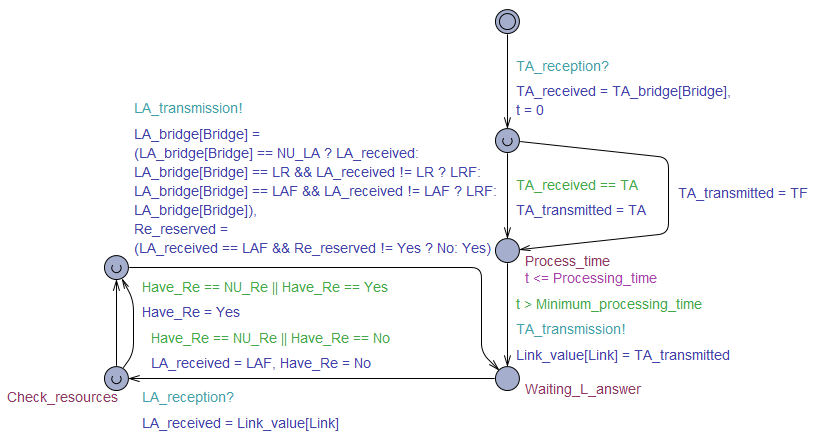}
\caption{BridgeOutput template.}
\label{fig:BOT}
\end{figure*}

Figures \ref{fig:BIT} and \ref{fig:BOT} depict the templates of the bridges. The first template performs the actions carried out by a port of the bridge when it receives a talker attribute and when it transmits the listener response. The second template performs the actions carried out by a port of the bridge when it forwards the talker attribute and when it receives a listener response. In this sense, each bridge will be composed of as many BridgeInput templates as ports through which talker attributes can be received and as many BridgeOutput templates as ports through which talker attributes should be forwarded.

In the bridges everything starts with the BridgeInput template receiving a talker attribute and forwarding it to all BridgeOutput templates of the bridge. After that the BridgeInput template will get stuck in the Waiting\_P\_answer location until it receives an answer from any BridgeOutput while the BridgeOutput templates takes the first edge. After that, the BridgesOutput templates will check their resources and, after the process time bounded between 10 and 200 ms already explained in previous section, each BridgeOutput template will transmit a \gls{TA} or \gls{TF} message through its corresponding port and will get stuck in the Waiting\_L\_answer location until they receive an answer from the device connected to the port, which can be a listener or another bridge.

When an answer is received, the BridgeOutput template of the corresponding port moves to the Check\_resources location while the BridgeInput location moves to the Process\_time location. During the process time the BridgeOutput template check the resources and determine the listener attribute to be transmitted. Finally, the BridgeOutput template return to the Waiting\_L\_answer location and the BridgeInput template transmit the answer through the port from which the bridge receives the talker attribute and return to the Waiting\_P\_answer location too.

Note that, if any other listener response arrives during the process time, it will be taken into account in the result because it is processed in 0 time units, so its contribution will not be lost; while, if the listener response arrives after the process time, another listener forwarding will be done in the bridge. This behaviour is the one expected from \gls{AVB} switches.


\section{Evaluation of the Termination of \gls{SRP}}\label{TE}



In this work we differentiate two levels of termination: termination for the application and for the infrastructure. The first one affects the nodes and, therefore, the application. The lack of termination at the application level can cause malfunction of some applications. This is due to the fact that many applications require to know the result of the reservation to make important decisions.

The infrastructure level refers to the bridges of the network. Even if in an ideal system these devices do not require termination, it is important to provide it to prevent unforeseen and undesirable effects in future reservations. For example, if a bridge receives many requests without resolution, it would be possible to cause an overflow of the buffer that could prevent the bridge from accepting new reservation requests or force it to eliminate some already accepted ones.

We next present the problems detected but it is important to note that the issues are mainly due to the fact that in \gls{SRP} listeners do not inform the bridges nor the talkers when they are not interested in binding to a stream.

\subsection{Termination at the Application Level}\label{TEAL}

Using the \uppaal model, we find a series of scenarios where the talker does not receive any response from the listeners and, thus, it waits indefinitely. This can happen, even in the absence of faults, when there are no listeners interested in the stream. As said before, many critical applications require to know the result of the reservations to make important decisions. Thus, the lack of termination can cause a malfunction of those applications, such as blocking the decision process or leading to incorrect decisions due to the lack of knowledge.

To check the termination for the application level we used the following queries:

\begin{equation}\label{equ:1}
\cmdtt{E[] T.LAs_received == NU_LA}
\end{equation}

\begin{equation} \label{equ:2}
\begin{split}
& \cmdtt{L0[].End && L1.End && L2.End &&}\\
& \cmdtt{(L0.LA_transmitted != NU_LA ||}\\
& \cmdtt{L1.LA_transmitted != NU_LA ||}\\
& \cmdtt{L2.LA_transmitted != NU_LA)}\\
& \cmdtt{--> T.LAs_received != NU_LA}
\end{split}
\end{equation}

\begin{equation} \label{equ:3}
\begin{split}
& \cmdtt{L0.End && L1.End && L2.End &&}\\
& \cmdtt{L0.LA_transmitted == NU_LA &&}\\
& \cmdtt{L1.LA_transmitted == NU_LA &&}\\
& \cmdtt{L2.LA_transmitted == NU_LA}\\
& \cmdtt{--> T.LAs_received == NU_LA}
\end{split}
\end{equation}

In the table included in the annex at the end of this document it is possible to see which queries are actually satisfied and which ones are not.

Query \ref{equ:1} checks if there is a path of states in the system (\cmdtt{E[]}) in which the talker does not receive any listener response (\cmdtt{T.LAs_received == NU_LA}). The query is satisfied so, it is possible that a talker does not receive any listener response. Then we checked if it is possible this to happen if at least one listener is interested in the stream. To do that we used the query \ref{equ:2}. This query checks if, at the end of the listeners actions (\cmdtt{L0.End && L1.End && L2.End}), at least one listener has replied something to the talker (\cmdtt{L0.LA_transmitted != NU_LA || L1.LA_transmitted != NU_LA || L2.LA_transmitted != NU_LA}), so at least one listener is interested in the stream, the talker receives at least one response (\cmdtt{T.LAs_received != NU_LA}). Finally, we checked that the non-reception of response by the talker was due to the non-transmission of response by the listeners. This was checked with the query \ref{equ:3}. This query checks if, at the end of the listeners actions (\cmdtt{L0.End && L1.End && L2.End}), no listeners has responded (\cmdtt{L0.LA_transmitted == NU_LA && L1.LA_transmitted == NU_LA && L2.LA_transmitted}), so there are no interested listeners in the stream, the talker receives no response (\cmdtt{T.LAs_received = NU_LA}).

\subsection{Termination at the Infrastructure Level}\label{TEIL}

A bridge that forwards the request of a talker waits for the responses of the listeners indefinitely. Also, bridges register talkers’ attributes in all their ports, and they do so for all the talkers willing to transmit. Similarly to what happens for termination at the application level, we find some scenarios where some bridges do not receive any response from the listeners, even in the absence of faults and even if the first level of termination is actually achieved by the protocol. Thus, bridges can wait indefinitely, e.g., if there are no listeners interested in the stream connected directly or indirectly to the bridge. This can cause an unnecessary use of memory in bridges and can later prevent the creation of streams with listeners willing to bind due to a lack of memory.

\begin{figure}[!t]
\centering
\includegraphics[width=0.5\columnwidth]{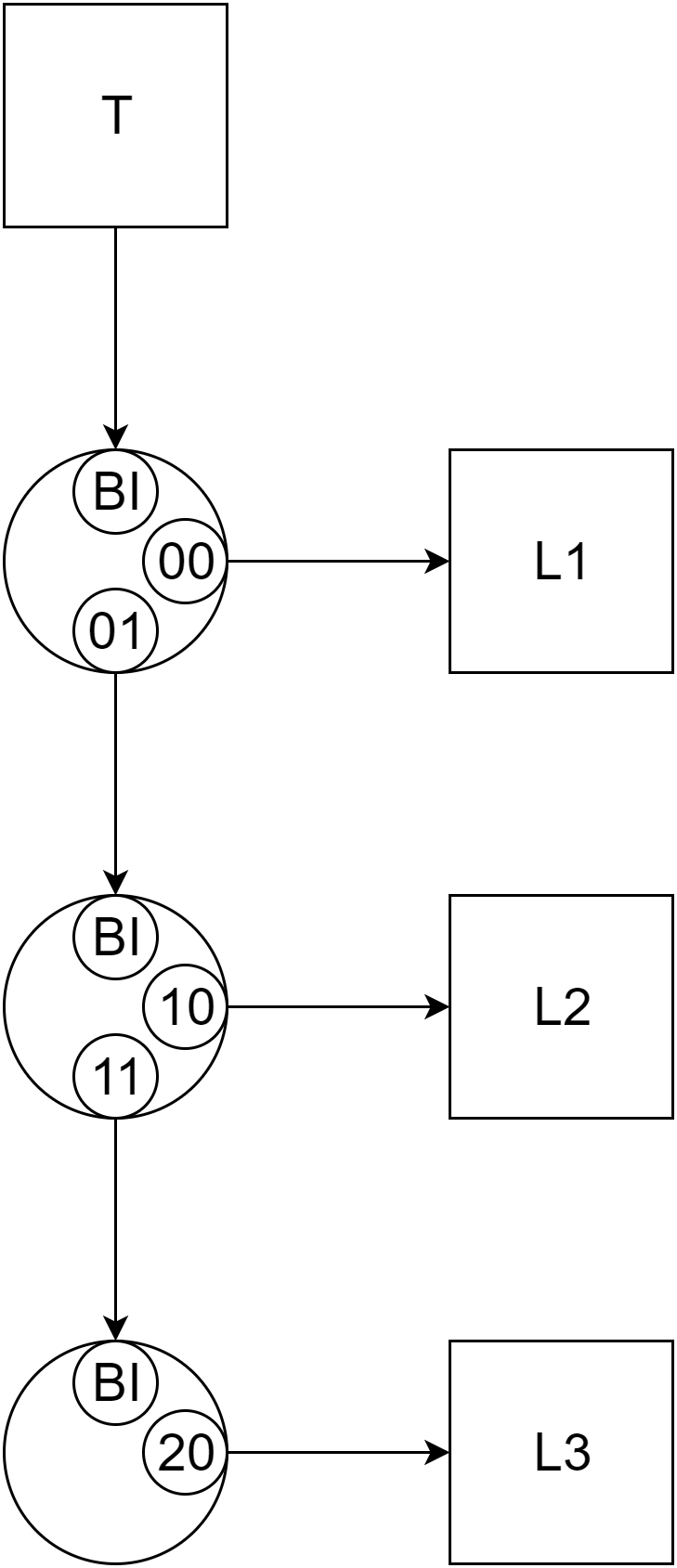}
\caption{BridgeOutput ports identification.}
\label{fig:BOP}
\end{figure}

To check the termination at the infrastructure level we used three different queries for each of the ports. These queries are similar to the ones used in the verification of the termination at the application level. Queries \ref{equ:4}, \ref{equ:7}, \ref{equ:10}, \ref{equ:13} and \ref{equ:16} check if exist a path of states in the system (\cmdtt{E[]}) in which the port of the bridge does not receive any listener response (\cmdtt{BQXY.LA_received == NU_LA}, where X is the bridge and Y is the identifier of the port as can be seen in Figure \ref{fig:BOP}). As the query is satisfied, it is possible that the port of the bridge does not receive any listener response. Then we checked if it is possible this to happen if at least one listener connected directly or indirectly to the bridge is interested in the stream. To do that we used queries \ref{equ:5}, \ref{equ:8}, \ref{equ:11}, \ref{equ:14} and \ref{equ:17}. These queries check if, at the end of the listeners actions (\cmdtt{LX.End && LY.End && ... && LN.End}, where the letters X, Y, N are the identifiers of the listeners connected directly or indirectly to the port of the bridge), at least one listener has replied something (\cmdtt{LX.LA_transmitted != NU_LA || LY.LA_transmitted != NU_LA || ... || LN.LA_transmitted != NU_LA}), so at least one listener is interested in the stream, the port of the bridge receives at least one response (\cmdtt{BQXY.LA_received != NU_LA}). Finally, we checked that the non-reception of response by the port of a bridge was due to the non-transmission of response by the listeners connected directly or indirectly to the port of the bridge. This was checked with the third queries \ref{equ:6}, \ref{equ:9}, \ref{equ:12}, \ref{equ:15} and \ref{equ:18}. These queries check if, at the end of the listeners actions (\cmdtt{LX.End && LY.End && ... && LN.End}), no listener has responded (\cmdtt{LX.LA_transmitted == NU_LA && LY.LA_transmitted == NU_LA && ... && LN.LA_transmitted}), so there are no interested listeners in the stream, the port of the bridge receives no response (\cmdtt{BQXY.LA_received == NU_LA}).

Queries of port 00:

\begin{equation} \label{equ:4}
\cmdtt{E[] BQ00.LA_received == NU_LA}
\end{equation}

\begin{equation} \label{equ:5}
\begin{split}
\cmdtt{L0.End && L0.LA_transmitted != NU_LA}\\
\cmdtt{--> BQ00.LA_received != NU_LA}
\end{split}
\end{equation}

\begin{equation} \label{equ:6}
\begin{split}
\cmdtt{L0.End && L0.LA_transmitted == NU_LA}\\
\cmdtt{--> BQ00.LA_received == NU_LA}
\end{split}
\end{equation}

Queries of port 01:

\begin{equation} \label{equ:7}
\cmdtt{E[] BQ01.LA_received == NU_LA}
\end{equation}

\begin{equation} \label{equ:8}
\begin{split}
& \cmdtt{L1.End && L2.End &&}\\
& \cmdtt{(L1.LA_transmitted != NU_LA ||}\\
& \cmdtt{L2.LA_transmitted != NU_LA)}\\
& \cmdtt{--> BQ01.LA_received != NU_LA}
\end{split}
\end{equation}

\begin{equation} \label{equ:9}
\begin{split}
& \cmdtt{L1.End && L2.End &&}\\
& \cmdtt{L1.LA_transmitted == NU_LA &&}\\
& \cmdtt{L2.LA_transmitted == NU_LA}\\
& \cmdtt{--> BQ01.LA_received == NU_LA}
\end{split}
\end{equation}

Queries of port 10:

\begin{equation} \label{equ:10}
\cmdtt{E[] BQ10.LA_received == NU_LA}
\end{equation}

\begin{equation} \label{equ:11}
\begin{split}
\cmdtt{L1.End && L1.LA_transmitted != NU_LA}\\
\cmdtt{--> BQ10.LA_received != NU_LA}
\end{split}
\end{equation}

\begin{equation} \label{equ:12}
\begin{split}
\cmdtt{L1.End && L1.LA_transmitted == NU_LA}\\
\cmdtt{--> BQ10.LA_received == NU_LA}
\end{split}
\end{equation}

Queries of port 11:

\begin{equation} \label{equ:13}
\cmdtt{E[] BQ11.LA_received == NU_LA}
\end{equation}

\begin{equation} \label{equ:14}
\begin{split}
\cmdtt{L2.End && L2.LA_transmitted != NU_LA}\\
\cmdtt{--> BQ11.LA_received != NU_LA}
\end{split}
\end{equation}

\begin{equation} \label{equ:15}
\begin{split}
\cmdtt{L2.End && L2.LA_transmitted == NU_LA}\\
\cmdtt{--> BQ11.LA_received == NU_LA}
\end{split}
\end{equation}

Queries of port 20:

\begin{equation} \label{equ:16}
\cmdtt{E[] BQ20.LA_received == NU_LA}
\end{equation}

\begin{equation} \label{equ:17}
\begin{split}
\cmdtt{L2.End && L2.LA_transmitted != NU_LA}\\
\cmdtt{--> BQ20.LA_received != NU_LA}
\end{split}
\end{equation}

\begin{equation} \label{equ:18}
\begin{split}
\cmdtt{L2.End && L2.LA_transmitted == NU_LA}\\
\cmdtt{--> BQ20.LA_received == NU_LA}
\end{split}
\end{equation}

In the table included in the annex at the end of this document it is possible to see which queries are actually satisfied and which ones are not.


\section{Evaluation of the Consistency of \gls{SRP}}\label{CE}



As in the previous section, we differentiate two levels of consistency: consistency for the application level and for the infrastructure level. Again, the first one affects the nodes and, therefore, the application. The lack of consistency at the application level can cause malfunction of some applications. Some applications require the different nodes to carry out coordinated actions because, e.g., they may rely on active replication of the nodes. In these applications, consistency in the communications is key to guarantee the correct operation of the overall system. The first step towards achieving consistent communications is to reserve the network resources consistently. Thus, at this level, \gls{SRP} should guarantee that enough listeners have resources reserved for the communication before starting to transmit.

As before, the infrastructure level refers to the bridges of the network. As we will see later, inconsistencies when reserving resources in bridges can cause waste of resources. This, in the long term, causes that streams, for which there would be sufficient resources, cannot be declared due to the resources reserved and wasted in some bridges.

As in the evaluation of the termination, despite the importance of consistency, we found some issues in both levels even in the absence of faults. We next present the problems detected but it is important to note that the issues are mainly due to the fact that information related to the reservations is propagated in a single direction. That is, the talker attribute transmitted by a talker is forwarded always towards the listeners; while, when listeners and bridges reply to a stream declaration, the information is only forwarded towards the talker. Thus, not all devices involved in the reservation of a stream receive the same information. We next describe the consistency issues detected and their effects.

\subsection{Consistency at the Application Level}\label{CEAL}

In \gls{SRP}, resources can be reserved for a subset of listeners, even when there are listeners willing to communicate that do not have resources to do it. In this case, the talker only communicates to a subset of listeners, generating an unnoticed inconsistency in the exchange of data. This means that actually starting a stream (with some listeners) has priority over doing it consistently (with either all or none of them). In addition, talkers cannot know which listeners have enough resources and which ones do not. A talker only knows if all interested listeners have enough resources when it receives \gls{LR} messages; if all interested listeners have not enough resources when it receives \gls{LAF} messages; if no listener is interested when it does not receive any answer; or, if at least one interested listener has enough resources when it receives \gls{LRF} messages. This limited information does not allow the talker to take intelligent decisions. Furthermore, we have to take into account that this information can change during the execution of the \gls{SRP} mechanism e.g. it is possible for a talker to receive an \gls{LR} message and then receive an \gls{LRF} message. Something similar can happen in listeners. They may be interested in the stream and have sufficient resources, but they do not receive anything because during the transmission of the response, the route to the talker did not have enough resources.

Furthermore, even when all listeners willing to bind have enough resources to do so, there are scenarios where consistency for the application is not guaranteed all the time. This can happen for two reasons, first the paths between a talker and different listeners may differ in length and end-to-end delay and, second, the talker starts transmitting as soon as it receives the response of one listener ready to receive. Therefore, some listeners willing to bind to the stream, with enough resources throughout the whole path towards the talker, may miss the first frames transmitted by the talker. This can cause, for example, two replicated nodes to be in two different states so that, although from that moment they receive the same data, they will not provide the same result.

To check the consistency for the application level we used the following queries:

\begin{equation} \label{equ:19}
\begin{split}
\cmdtt{E<> }& \cmdtt{S.Stream_transmission &&}\\
& \cmdtt{L0.LA_transmitted == LR &&}\\
& \cmdtt{BQ00.Re_reserved == No}
\end{split}
\end{equation}

\begin{equation} \label{equ:20}
\begin{split}
\cmdtt{E<> }& \cmdtt{S.Stream_transmission &&}\\
& \cmdtt{L1.LA_transmitted == LR &&}\\
& \cmdtt{(BQ01.Re_reserved == No ||}\\
& \cmdtt{BQ10.Re_reserved == No)}
\end{split}
\end{equation}

\begin{equation} \label{equ:21}
\begin{split}
\cmdtt{E<> }& \cmdtt{S.Stream_transmission &&}\\
& \cmdtt{L2.LA_transmitted == LR &&}\\
& \cmdtt{(BQ01.Re_reserved == No ||}\\
& \cmdtt{BQ11.Re_reserved == No ||}\\
& \cmdtt{BQ20.Re_reserved == No)}
\end{split}
\end{equation}

\begin{equation} \label{equ:22}
\begin{split}
& \cmdtt{S.Stream_transmission &&}\\
& \cmdtt{L2.LA_transmitted == LR &&}\\
& \cmdtt{(BQ01.Re_reserved == NU_Re ||}\\
& \cmdtt{BQ11.Re_reserved == NU_Re ||}\\
& \cmdtt{BQ20.Re_reserved == NU_Re)}\\
& \cmdtt{--> !(BQ01.Re_reserved == Yes &&}\\
& \cmdtt{BQ11.Re_reserved == Yes &&}\\
& \cmdtt{BQ20.Re_reserved == Yes) &&}\\
& \cmdtt{(BQ01.Re_reserved != NU_Re &&}\\
& \cmdtt{BQ11.Re_reserved != NU_Re &&}\\
& \cmdtt{BQ20.Re_reserved != NU_Re)}
\end{split}
\end{equation}

In the table included in the annex at the end of this document it is possible to see which queries are actually satisfied and which ones are not.

Queries \ref{equ:19}, \ref{equ:20} and \ref{equ:21} check if there is at least one state in which the talker is already transmitting (\cmdtt{S.Stream_transmission}), a listener is interested in the stream and, from his point of view, has sufficient resources (\cmdtt{LX.LA_transmitted == LR} where X is the identifier of the listener) but the route from the talker to the listener has not reserved the necessary resources for that stream (e.g. \cmdtt{BQ01.Re_reserved == No || BQ11.Re_reserved == No || BQ20.Re_reserved == No} for listener L2). These tests show that the talker can start transmitting even when there are interested listeners that will not be able to receive the stream. Moreover, it also shows that there are listeners that believe they will receive the stream but never will.

Query \ref{equ:22} was used to verify that even when all interested listeners can bind to the stream some of them may miss the first messages because the talker starts transmitting before finishing the resource reservation. Specifically, it checks if a talker transmitting (\cmdtt{S.Stream_transmission}), a listener waiting for the stream (\cmdtt{L2.LA_transmitted == LR}) and the route not yet reserved (\cmdtt{BQ01.Re_reserved == NU_Re || BQ11.Re_reserved == NU_Re || BQ20.Re_reserved == NU_Re}) implies that the route will never be reserved (\cmdtt{!(BQ01.Re_reserved == Yes && BQ11.Re_reserved == Yes && BQ20.Re_reserved == Yes) && (BQ01.Re_reserved != NU_Re && BQ11.Re_reserved != NU_Re && BQ20.Re_reserved != NU_Re)}). As the query is not satisfied we proved the inconsistency in the data received at the beginning of the stream.

\subsection{Consistency at the Infrastructure Level}\label{CEIL}

In this work we also find out that bridges can make inconsistent decisions regarding the reservation of resources of a stream. Specifically, in \gls{SRP} it is possible that some bridges reserve resources for a stream but other bridges in the same route to the listener do not. This implies a waste of resources in the bridges that reserved the resources because the listeners for which they reserved the resources are not going to receive the stream because of the bridges in the same route that did not reserve the resources. This may not be problematic at first, but, with an utilisation close to 100\%, this may cause streams, for which there would be sufficient resources, unable be declared due to the resources wasted in these bridges.

To check the consistency for the infrastructure level we used the following queries:

\begin{equation} \label{equ:23}
\begin{split}
\cmdtt{E<> }& \cmdtt{deadlock &&}\\
& \cmdtt{S.Stream_transmission &&}\\
& \cmdtt{BQ01.Re_reserved == No &&}\\
& \cmdtt{BQ10.Re_reserved == Yes &&}\\
& \cmdtt{BQ11.Re_reserved == Yes &&}\\
& \cmdtt{BQ20.Re_reserved == Yes}
\end{split}
\end{equation}

\begin{equation} \label{equ:24}
\begin{split}
\cmdtt{E<> }& \cmdtt{deadlock &&}\\
& \cmdtt{S.Stream_transmission &&}\\
& \cmdtt{(BQ01.Re_reserved == No ||}\\
& \cmdtt{BQ11.Re_reserved == No) &&}\\
& \cmdtt{BQ20.Re_reserved == Yes}
\end{split}
\end{equation}

\begin{equation} \label{equ:25}
\begin{split}
\cmdtt{S.Stream_transmission}\\
\cmdtt{--> (BQ00.Re_reserved == Yes &&}\\
\cmdtt{BQ01.Re_reserved != Yes &&}\\
\cmdtt{BQ10.Re_reserved != Yes &&}\\
\cmdtt{BQ11.Re_reserved != Yes &&}\\
\cmdtt{BQ20.Re_reserved != Yes &&}\\
\cmdtt{L0.LA_transmitted == LR &&}\\
\cmdtt{L1.LA_transmitted != LR &&}\\
\cmdtt{L2.LA_transmitted != LR) ||}\\
\cmdtt{(BQ00.Re_reserved != Yes &&}\\
\cmdtt{BQ01.Re_reserved == Yes &&}\\
\cmdtt{BQ10.Re_reserved == Yes &&}\\
\cmdtt{BQ11.Re_reserved != Yes &&}\\
\cmdtt{BQ20.Re_reserved != Yes &&}\\
\cmdtt{L0.LA_transmitted != LR &&}\\
\cmdtt{L1.LA_transmitted == LR &&}\\
\cmdtt{L2.LA_transmitted != LR) ||}\\
\cmdtt{(BQ00.Re_reserved != Yes &&}\\
\cmdtt{BQ01.Re_reserved == Yes &&}\\
\cmdtt{BQ10.Re_reserved != Yes &&}\\
\cmdtt{BQ11.Re_reserved == Yes &&}\\
\cmdtt{BQ20.Re_reserved == Yes &&}\\
\cmdtt{L0.LA_transmitted != LR &&}\\
\cmdtt{L1.LA_transmitted != LR &&}\\
\cmdtt{L2.LA_transmitted == LR) ||}\\
\cmdtt{(BQ00.Re_reserved == Yes &&}\\
\cmdtt{BQ01.Re_reserved == Yes &&}\\
\cmdtt{BQ10.Re_reserved == Yes &&}\\
\cmdtt{BQ11.Re_reserved != Yes &&}\\
\cmdtt{BQ20.Re_reserved != Yes &&}\\
\cmdtt{L0.LA_transmitted == LR &&}\\
\cmdtt{L1.LA_transmitted == LR &&}\\
\cmdtt{L2.LA_transmitted != LR)}
\end{split}
\end{equation}

In the table included in the annex at the end of this document it is possible to see which queries are actually satisfied and which ones are not.

Query \ref{equ:23} checks if there is at least one state (\cmdtt{E<>}) after the mechanism has been executed (\cmdtt{deadlock}) in which the stream is being transmitted (\cmdtt{S.Stream_transmission}) while the link that supplies the bridges 1 and 2 is not reserved but the links of the bridges 1 and 2 are. This reservation distribution implies a waste of resources in all the links reserved by the bridges 1 and 2 because, as the link that supplies them is not reserved, they are not going to receive data messages from this stream. Query \ref{equ:24} is almost the same but it checks the scenario where only the bridge 2 is being affected by the inconsistency issue. Finally, query \ref{equ:25} checks if the transmission of the stream (\cmdtt{S.Stream_transmission}) always implies one of all correct distributions of resource reservations. As it is not satisfied, we can determine that incorrect distributions of resource reservations (with waste of resources) can be achieved by the protocol.


\section{CSRP Description}\label{SP}

\gls{CSRP}, as \gls{SRP}, follows the publisher-subscriber paradigm, where the publisher is called talker and the subscribers, listeners. The real-time data communications are made through streams, a logical communication channel that carries traffic defined by a set of parameters, such as the period or frame size.

To create a stream the talker must declare its intention to communicate by transmitting a \gls{TA} message, which is still in broadcast mode. The \gls{TA} message contains information to identify the stream and the resources it needs. The bridges process the message and check if there are enough resources in the forwarding ports to create the stream. If there are enough resources in a port, the bridge forwards the \gls{TA} message through it; otherwise, if the port does not have sufficient resources, the bridge transmits a \gls{TF} message through it. A \gls{TF} message is also transmitted in broadcast mode and contains the same information as a \gls{TA} message but adding the reason why the resource reservation has failed. At this point, as in the standardised version of \gls{SRP}, the bridges record the talker's request but do not yet make the reservation of resources.

When a listener receives a talker attribute, it decides if he wants to join the stream or not. If the listener is not interested in the stream, it will not take any action or inform anyone about its decision. In contrast, if the listener is interested in the stream, 3 different scenarios can happen: (i) if the listener receives a \gls{TA} message, the listener checks its resources and, if it has enough to receive the stream, transmits an \gls{LR} message; (ii) if when checking their resources these are not enough, the listener will transmit an \gls{LAF} message and (iii) if the listener receives a \gls{TA} message it will also transmit an \gls{LAF} message.

The first modification of the protocol is found in the transmission of listener attributes by the bridges. Bridges receive the listener attributes and combine them to send them to the talker. In order to accomplish this, bridges analyse the responses received by each port and then generate the new response that they transmit towards the talker. Whenever a bridge receives an \gls{LR} message through a port, it checks whether the port has enough resources. If there are enough resources, the \gls{LR} remains unchanged and the port reserves the necessary resources provisionally, instead of definitely like in \gls{SRP}; otherwise, the \gls{LR} becomes an \gls{LAF} message. On the other hand, if the bridge receives an \gls{LAF} message the value is left unchanged and the port does not reserve the resources. In case of concurrent requests, and this is another change with respect to \gls{SRP}, the provisional reservation is made for the first \gls{LR} or \gls{LRF} message received, while the rest are transferred to a \gls{FIFO} list. The items in this list are only deleted when their reservation processes are completed or when the reservation of resources is confirmed.

After processing the listener attributes, each bridge must join them to forward an updated one to the talker. Whenever a bridge has several listener responses to forward, it combines the responses into a single one and transmits it to the talker. The result of combining the responses is the following: (i) if the bridge receives an \gls{LR} in all the ports, it transmits to the talker an \gls{LR} message; if the bridge receives an \gls{LAF} in all the ports, it transmits to the talker another \gls{LAF} message; and, if the bridge receives \gls{LR} messages in some ports and \gls{LAF} messages in other ports, it will transmits to the talker a new message called \gls{LRF} message. It is important to note that the bridges do not wait for the reception of all the listener attributes, but they are continuously joining and retransmitting them as they receive new answers. In this way a bridge can transmit an \gls{LR} or \gls{LAF} message and then transmit an \gls{LRF} message, just like in \gls{SRP}. Nevertheless, in \gls{CSRP} bridges must specify in the listener attribute which listeners can receive and which listeners cannot. To do so, \gls{CSRP} relies on two lists, one for successful reservations and one for unsuccessful ones. Specifically, edge bridges introduce the identifier of the node that sends the \gls{LR} or \gls{LAF} message in the corresponding list and sends them embedded in the response to the talker. Whenever a bridge receives a response from another bridge, it checks the lists and updates them accordingly when joining the responses.

The talker waits for the answers for a bounded period of time, determined by a local timer that the talker activates when transmitting the \gls{TA}. After that time, the talker uses the lists with the node identifiers to know which listeners can receive and which listeners cannot and it decides whether to transmit the stream to all the listeners that can receive, to a subgroup or to none of them. This decision is communicated by transmitting in broadcast mode a message called \gls{FD}, which contains a list of listeners that will receive the stream and listeners that will not receive the stream.

When a bridge receives the \gls{FD} message it knows which listeners are going to receive the stream and which are not. In this way, bridges can lock the resources or eliminate unnecessary reservations. Listeners, on the other hand, can know whether they are subscribed to the stream or not so they do not wait indefinitely for the data transmission.

Once the \gls{FD} message has been transmitted and the resource reservation mechanism has finished, the talker starts transmitting the data stream. Finally, as in standard \gls{SRP}, once the stream has been created, the talker can delete it at any time by means of the unadvertised stream mechanism.


\section{\gls{CSRP} \uppaal Model}\label{CSRP_model}

The \uppaal model of \gls{CSRP} has the same topology, same templates, same instantiations of the templates and same abstractions as the model of the standardised \gls{SRP} explained in Section \ref{U_SRP_M}. To formally verify the correction of the improved mechanism (\gls{CSRP}'s resource reservation mechanism), we modified as little as possible the model shown above to include the changes proposed in our solution.

In the Talker template we basically eliminated the instantaneous transmission of data that occurred as soon as the speaker received an \gls{LR} or \gls{LRF} message. On the other hand, we added a timer to define the waiting time for listener responses and implemented the transmission of the \gls{FD} message.

In the bridge templates we implemented the reception and forwarding of the \gls{FD} message and the mechanisms to change the resource reservations based on it.

Finally, in the Listener template we implemented the reception of the \gls{FD} message and the mechanism so that listeners know if they can receive or not. It is important to remember that these modifications in listeners are not essential. However, not implementing them would imply that listeners remain unsure of whether they will receive or not until they receive any data message of the stream.

\subsection{Talker Template}

\begin{figure*}[!t]
\centering
\includegraphics[width=0.8\textwidth]{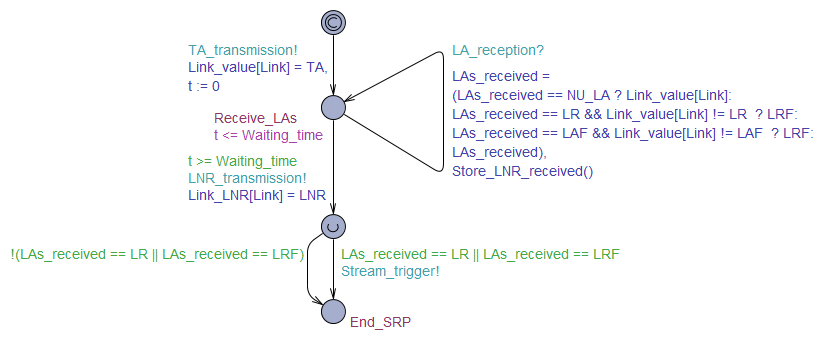}
\caption{Talker template with the proposed solution applied.}
\label{fig:TTS}
\end{figure*}

Figures \ref{fig:TTS} and \ref{fig:ST} show the talker templates. The first and most complex is the one that performs the main actions of \gls{SRP} in the talker whose differences with respect to the standard talker template we will discuss next. The other template, which represents the transmission of data frames, is the same as the previous model.

The main difference between the talker template shown in Figure \ref{fig:TTS} and the previous talker template shown in Figure \ref{fig:TT} is that in the proposed solution the left edge has been replaced by a path that end in the End\_SRP location. The left edge of the previous model activate the data transmission as soon as the talker receives an \gls{LR} or \gls{LRF} message. However, the model with the proposed solution has a path, which is activated by a timer, that transmits the \gls{FD} message and, if the talker has received an \gls{LR} or \gls{LRF} message during the waiting time, the talker will also activate the data transmission. This is a simple condition to start the transmission. However, the new information present in the talker would allow it to make much more complex decisions.

\subsection{Listener Template}

\begin{figure}[!t]
\centering
\includegraphics[width=0.7\columnwidth]{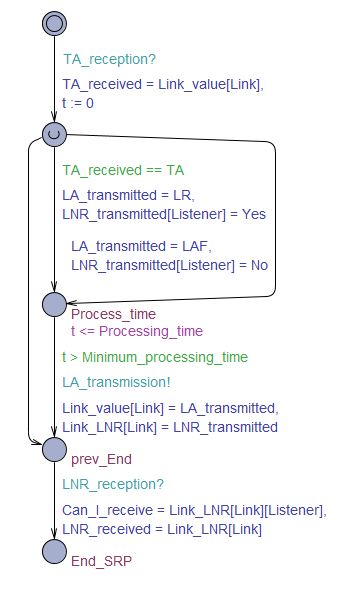}
\caption{Listener template with the proposed solution applied.}
\label{fig:LTS}
\end{figure}

Figure \ref{fig:LTS} shows the listener template with the proposed solution implemented. The first main difference between this template and the previous one (Figure \ref{fig:LT}) is that this one has some variables called \gls{LNR} which are used to convey the ID of the listener in the listener response. In addition, after the location prev\_End (location End in the previous listener template) in this template there is another edge and location. These receive the \gls{FD} message and store the final configuration of the resource reservation.

\subsection{Bridge Templates}

\begin{figure*}[!t]
\centering
\includegraphics[width=0.6\textwidth]{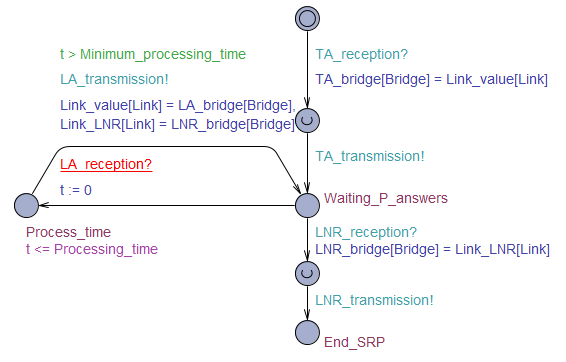}
\caption{BridgeInput template with the proposed solution applied.}
\label{fig:BITS}
\end{figure*}

\begin{figure*}[!t]
\centering
\includegraphics[width=0.8\textwidth]{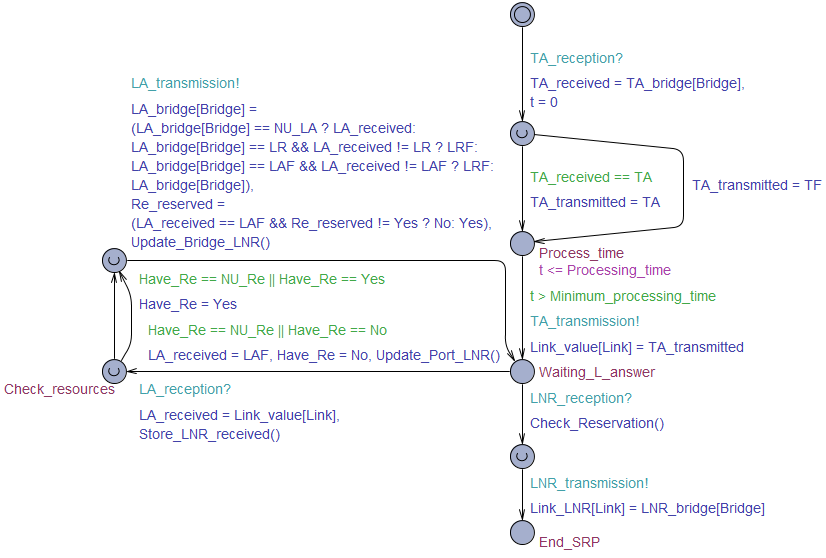}
\caption{BridgeOutput template with the proposed solution applied.}
\label{fig:BOTS}
\end{figure*}

Figures \ref{fig:BITS} and \ref{fig:BOTS} show the bridge templates with the proposed solution implemented. The main difference between these templates and the previous ones are: (i) these templates uses \gls{LNR} variables to share the status of resource reservation and (ii) the presence of a path after the Waiting\_X\_answer locations which end at the End\_SRP location. This path receives and forwards the \gls{FD} message while update the status of resource reservation in the bridge.


\section{Evaluation of the Termination of \gls{CSRP}}\label{CSRP_TE}

We next describe the validation of \gls{CSRP} from the termination point of view. Again, we address the issues at the application and infrastructure level. To do that, we used the same queries that proved the non-termination in \gls{SRP} plus some additional queries.

Just like in \gls{SRP}, if in \gls{CSRP} no nodes want to bind to a stream, the talker and bridges do not receive any listener response. Thus, we provide termination with the timer in the talker and the \gls{FD} message in the bridges, so now both, talkers and bridges, know when to stop waiting for listener responses.

\subsection{Termination at the Application Level}

In \gls{CSRP} it is still possible that the talker does not receive any response from the listeners (see Sub-section \ref{TEAL}). However, using the timer, the talker always stops waiting for an answer, makes a decision based on the information it has received and informs about it, by means of the \gls{FD} message, the network. We used the \uppaal model of \gls{CSRP} to validate the behaviour of the protocol. Specifically, we check that all the nodes finish the resource reservation process within a bounded time determined by the timer in the talker and the distance between the talker and the listeners.

To check the termination for the application level we used the following queries:

\begin{equation} \label{equ:26}
\begin{split}
\cmdtt{E[] T.LAs_received == NU_LA}
\end{split}
\end{equation}

\begin{equation} \label{equ:27}
\begin{split}
& \cmdtt{L0.prev_End &&}\\
& \cmdtt{L1.prev_End &&}\\
& \cmdtt{L2.prev_End &&}\\
& \cmdtt{(L0.LA_transmitted != NU_LA ||}\\
& \cmdtt{L1.LA_transmitted != NU_LA ||}\\
& \cmdtt{L2.LA_transmitted != NU_LA)}\\
& \cmdtt{--> T.LAs_received != NU_LA}
\end{split}
\end{equation}

\begin{equation} \label{equ:28}
\begin{split}
& \cmdtt{L0.prev_End &&}\\
& \cmdtt{L1.prev_End &&}\\
& \cmdtt{L2.prev_End &&}\\
& \cmdtt{L0.LA_transmitted != NU_LA &&}\\
& \cmdtt{L1.LA_transmitted != NU_LA &&}\\
& \cmdtt{L2.LA_transmitted != NU_LA}\\
& \cmdtt{--> T.LAs_received != NU_LA}
\end{split}
\end{equation}

\begin{equation} \label{equ:29}
\begin{split}
\cmdtt{A<> T.End_SRP}
\end{split}
\end{equation}

In the table included in the annex at the end of this document it is possible to see which queries are actually satisfied and which ones are not.

As we can see in queries \ref{equ:26}, \ref{equ:27} and \ref{equ:28}, also used in the previous model, in \gls{CSRP} it is still possible that the talker does not receive any response from the listeners. However, as we can see in query \ref{equ:29}, thanks to the timer, the talker always (\cmdtt{A<>}) stops waiting for an answer, makes a decision based on the received information and informs about it by means of the \gls{FD} message to the other devices of the network (\cmdtt{T.End_SRP}).

\subsection{Termination at the Infrastructure Level}

With this evaluation we see how bridges' ports may not receive any listener response. However, thanks to the \gls{FD} message sent by the talker, bridges always stop waiting for an answer and change their reserved resources based on the talker decision. We used the \uppaal model of \gls{SRP} to verify that all bridges finish the resource reservation process within a bounded time.

To check the termination for the infrastructure level we used the following queries:

Queries of port 00:

\begin{equation} \label{equ:30}
\cmdtt{E[] BQ00.LA_received == NU_LA}
\end{equation}

\begin{equation} \label{equ:31}
\begin{split}
& \cmdtt{L0.prev_End &&}\\
& \cmdtt{L0.LA_transmitted != NU_LA}\\
& \cmdtt{--> BQ00.LA_received != NU_LA}
\end{split}
\end{equation}

\begin{equation} \label{equ:32}
\begin{split}
& \cmdtt{L0.prev_End &&}\\
& \cmdtt{L0.LA_transmitted == NU_LA}\\
& \cmdtt{--> BQ00.LA_received == NU_LA}
\end{split}
\end{equation}

\begin{equation} \label{equ:33}
\begin{split}
\cmdtt{A<> BQ00.End_SRP}
\end{split}
\end{equation}

Queries of port 01:

\begin{equation} \label{equ:34}
\cmdtt{E[] BQ01.LA_received == NU_LA}
\end{equation}

\begin{equation} \label{equ:35}
\begin{split}
& \cmdtt{L1.prev_End && L2.prev_End &&}\\
& \cmdtt{(L1.LA_transmitted != NU_LA ||}\\
& \cmdtt{L2.LA_transmitted != NU_LA)}\\
& \cmdtt{--> BQ01.LA_received != NU_LA}
\end{split}
\end{equation}

\begin{equation} \label{equ:36}
\begin{split}
& \cmdtt{L1.prev_End && L2.prev_End &&}\\
& \cmdtt{L1.LA_transmitted == NU_LA &&}\\
& \cmdtt{L2.LA_transmitted == NU_LA}\\
& \cmdtt{--> BQ01.LA_received == NU_LA}
\end{split}
\end{equation}

\begin{equation} \label{equ:37}
\begin{split}
\cmdtt{A<> BQ01.End_SRP}
\end{split}
\end{equation}

Queries of port 10:

\begin{equation} \label{equ:38}
\cmdtt{E[] BQ10.LA_received == NU_LA}
\end{equation}

\begin{equation} \label{equ:39}
\begin{split}
& \cmdtt{L1.prev_End &&}\\
& \cmdtt{L1.LA_transmitted != NU_LA}\\
& \cmdtt{--> BQ10.LA_received != NU_LA}
\end{split}
\end{equation}

\begin{equation} \label{equ:40}
\begin{split}
& \cmdtt{L1.prev_End &&}\\
& \cmdtt{L1.LA_transmitted == NU_LA}\\
& \cmdtt{--> BQ10.LA_received == NU_LA}
\end{split}
\end{equation}

\begin{equation} \label{equ:41}
\begin{split}
\cmdtt{A<> BQ10.End_SRP}
\end{split}
\end{equation}

Queries of port 11:

\begin{equation} \label{equ:42}
\cmdtt{E[] BQ11.LA_received == NU_LA}
\end{equation}

\begin{equation} \label{equ:43}
\begin{split}
& \cmdtt{L2.prev_End &&}\\
& \cmdtt{L2.LA_transmitted != NU_LA}\\
& \cmdtt{--> BQ11.LA_received != NU_LA}
\end{split}
\end{equation}

\begin{equation} \label{equ:44}
\begin{split}
& \cmdtt{L2.prev_End &&}\\
& \cmdtt{L2.LA_transmitted == NU_LA}\\
& \cmdtt{--> BQ11.LA_received == NU_LA}
\end{split}
\end{equation}

\begin{equation} \label{equ:45}
\begin{split}
\cmdtt{A<> BQ11.End_SRP}
\end{split}
\end{equation}

Queries of port 20:

\begin{equation} \label{equ:46}
\cmdtt{E[] BQ20.LA_received == NU_LA}
\end{equation}

\begin{equation} \label{equ:47}
\begin{split}
& \cmdtt{L2.prev_End &&}\\
& \cmdtt{L2.LA_transmitted != NU_LA}\\
& \cmdtt{--> BQ20.LA_received != NU_LA}
\end{split}
\end{equation}

\begin{equation} \label{equ:48}
\begin{split}
& \cmdtt{L2.prev_End &&}\\
& \cmdtt{L2.LA_transmitted == NU_LA}\\
& \cmdtt{--> BQ20.LA_received == NU_LA}
\end{split}
\end{equation}

\begin{equation} \label{equ:49}
\begin{split}
\cmdtt{A<> BQ20.End_SRP}
\end{split}
\end{equation}

In the table included in the annex at the end of this document it is possible to see which queries are actually satisfied and which ones are not.

Here, as in the previous sub-section, we can see how bridges' ports may not receive any listener response in the first three queries of each port (\ref{equ:30}, \ref{equ:31}, \ref{equ:32}, \ref{equ:34}, \ref{equ:35}, \ref{equ:36}, \ref{equ:38}, \ref{equ:39},  \ref{equ:40}, \ref{equ:42}, \ref{equ:43}, \ref{equ:44}, \ref{equ:46}, \ref{equ:47} and \ref{equ:48}), the ones used in the previous model. However, as we can see in the last query of each port (\ref{equ:33}, \ref{equ:37}, \ref{equ:41}, \ref{equ:45}, \ref{equ:49}), thanks to the \gls{FD} message, the bridges always (\cmdtt{A<>}) stop waiting for answers and change their reserved resources based on the talker decision (\cmdtt{BQXY.End_SRP} where X indicates the bridge and Y the forwarding port of the bridge).


\section{Evaluation of the Consistency of \gls{CSRP}}\label{CSRP_CE}

In this section we describe the validation of \gls{CSRP} from the consistency point of view, at the application and infrastructure level. To do so, we use the same queries that proved the inconsistency in \gls{SRP} plus some additional queries.
This solution solves all the detected consistency issues. We achieved this by centralising the decisions in the talker and ensuring the homogeneous propagation of information related to the reservation of resources.

\subsection{Consistency at the Application Level}

First, note that this solution does not aim at providing resources for all the listeners that want to bind. Instead, it aims at ensuring that all listeners know what is the status of the reservation regardless of whether they can receive or not. This was not guaranteed in the standard \gls{SRP} but it is achieved in \gls{CSRP} thanks to the \gls{FD} message. 
We verify the consistent view of the network. Specifically, we prove that when \gls{CSRP} finishes the reservation process, all devices know which nodes are subscribed to the stream and which are not, including the nodes.

To check the consistency for the application level we used the following queries:

\begin{equation} \label{equ:50}
\begin{split}
\cmdtt{E<> }& \cmdtt{S.Stream_transmission &&}\\
& \cmdtt{L0.LA_transmitted == LR &&}\\
& \cmdtt{BQ00.Re_reserved == No}
\end{split}
\end{equation}

\begin{equation} \label{equ:51}
\begin{split}
\cmdtt{E<> }& \cmdtt{S.Stream_transmission &&}\\
& \cmdtt{L0.L0.Can_I_receive == Yes &&}\\
& \cmdtt{BQ00.Re_reserved == No}
\end{split}
\end{equation}

\begin{equation} \label{equ:52}
\begin{split}
\cmdtt{A[] }& \cmdtt{L0.Can_I_receive == Yes imply}\\
& \cmdtt{BQ00.Re_reserved == Yes}
\end{split}
\end{equation}

\begin{equation} \label{equ:53}
\begin{split}
\cmdtt{E<> }& \cmdtt{S.Stream_transmission &&}\\
& \cmdtt{L1.LA_transmitted == LR &&}\\
& \cmdtt{(BQ01.Re_reserved == No ||}\\
& \cmdtt{BQ10.Re_reserved == No)}
\end{split}
\end{equation}

\begin{equation} \label{equ:54}
\begin{split}
\cmdtt{E<> }& \cmdtt{S.Stream_transmission &&}\\
& \cmdtt{L1.Can_I_receive == Yes &&}\\
& \cmdtt{(BQ01.Re_reserved == No ||}\\
& \cmdtt{BQ10.Re_reserved == No)}
\end{split}
\end{equation}

\begin{equation} \label{equ:55}
\begin{split}
\cmdtt{A[] }& \cmdtt{L1.Can_I_receive == Yes imply}\\
& \cmdtt{BQ01.Re_reserved == Yes &&}\\
& \cmdtt{BQ10.Re_reserved == Yes}
\end{split}
\end{equation}

\begin{equation} \label{equ:56}
\begin{split}
\cmdtt{E<> }& \cmdtt{S.Stream_transmission &&}\\
& \cmdtt{L2.LA_transmitted == LR &&}\\
& \cmdtt{(BQ01.Re_reserved == No ||}\\
& \cmdtt{BQ11.Re_reserved == No ||}\\
& \cmdtt{BQ20.Re_reserved == No)}
\end{split}
\end{equation}

\begin{equation} \label{equ:57}
\begin{split}
\cmdtt{E<> }& \cmdtt{S.Stream_transmission &&}\\
& \cmdtt{L2.Can_I_receive == Yes &&}\\
& \cmdtt{(BQ01.Re_reserved == No ||}\\
& \cmdtt{BQ11.Re_reserved == No ||}\\
& \cmdtt{BQ20.Re_reserved == No)}
\end{split}
\end{equation}

\begin{equation} \label{equ:58}
\begin{split}
\cmdtt{A[] }& \cmdtt{L2.Can_I_receive == Yes imply}\\
& \cmdtt{BQ01.Re_reserved == Yes &&}\\
& \cmdtt{BQ11.Re_reserved == Yes &&}\\
& \cmdtt{BQ20.Re_reserved == Yes}
\end{split}
\end{equation}

\begin{equation} \label{equ:59}
\begin{split}
\cmdtt{A[] }& \cmdtt{deadlock imply}\\
& \cmdtt{T.End_SRP && L0.End_SRP &&}\\
& \cmdtt{L1.End_SRP && L2.End_SRP &&}\\
& \cmdtt{BI0.End_SRP && BI1.End_SRP &&}\\
& \cmdtt{BI2.End_SRP && BQ00.End_SRP &&}\\
& \cmdtt{BQ01.End_SRP && BQ10.End_SRP &&}\\
& \cmdtt{BQ11.End_SRP && BQ20.End_SRP}
\end{split}
\end{equation}

\begin{equation} \label{equ:60}
\begin{split}
\cmdtt{A[] }& \cmdtt{deadlock imply}\\
& \cmdtt{T.LNR == LNR_bridge[0] &&}\\
& \cmdtt{T.LNR == LNR_bridge[1] &&}\\
& \cmdtt{T.LNR == LNR_bridge[2] &&}\\
& \cmdtt{T.LNR == L0.LNR_received &&}\\
& \cmdtt{T.LNR == L1.LNR_received &&}\\
& \cmdtt{T.LNR == L2.LNR_received}
\end{split}
\end{equation}

In the table included in the annex at the end of this document it is possible to see which queries are actually satisfied and which ones are not.

Queries \ref{equ:50}, \ref{equ:53}, \ref{equ:56} show that there are states (\cmdtt{E<>}) where a listener wants to bind to the stream and it thinks it can (\cmdtt{L2.LA_transmitted == LR}) but the resources have not been reserved (\cmdtt{BQ01.Re_reserved == No || BQ11.Re_reserved == No || BQ20.Re_reserved == No}). However, thanks to the \gls{FD} message, as we can see in queries \ref{equ:51}, \ref{equ:52}, \ref{equ:54}, \ref{equ:55}, \ref{equ:57}, \ref{equ:58}, now listeners know when they can and when they cannot receive the stream.

Finally, queries \ref{equ:59} and \ref{equ:60} verify the consistent view of the network. Query \ref{equ:59} verifies that always (\cmdtt{A[]}) a deadlock implies the end of the reservation process (\cmdtt{T.End_SRP} \cmdtt{&& L0.End_SRP && L1.End_SRP && L2.End_SRP && BI0.End_SRP && BI1.End_SRP && BI2.End_SRP && BQ00.End_SRP && BQ01.End_SRP && BQ10.End_SRP && BQ11.End_SRP && BQ20.End_SRP}) while query \ref{equ:60} checks that always (\cmdtt{A[]}) at the end of the reservation process (\cmdtt{deadlock}) all the \gls{LNR} are consistent (\cmdtt{T.LNR == LNR_bridge[0] && T.LNR == LNR_bridge[1] && T.LNR == LNR_bridge[2] && T.LNR == L0.LNR_received && T.LNR == L1.LNR_received && T.LNR == L2.LNR_received}).

\subsection{Consistency at the Infrastructure Level}

Finally, at the infrastructure level, we verify that \gls{CSRP} avoids wasting resources with unnecessary reservations thanks to the \gls{FD} message that informs the bridges about which listeners can bind to the stream and which listeners cannot. In this way, the bridges can free the resources they reserved for the listeners that cannot receive. We carry out this verification using the \gls{CSRP} \uppaal model.

To check the consistency for the infrastructure level we used queries \ref{equ:61}, \ref{equ:62} and \ref{equ:63}. At the infrastructure level, we not only avoid wasting resources with unnecessary reservations, as can be seen in queries \ref{equ:61} and \ref{equ:62}, which were satisfied for the previous model, but we also made sure that only the appropriate reservation distributions could be generated \ref{equ:63}.

\begin{equation} \label{equ:61}
\begin{split}
\cmdtt{E<> }& \cmdtt{deadlock &&}\\
& \cmdtt{S.Stream_transmission &&}\\
& \cmdtt{BQ01.Re_reserved == No &&}\\
& \cmdtt{BQ10.Re_reserved == Yes &&}\\
& \cmdtt{BQ11.Re_reserved == Yes &&}\\
& \cmdtt{BQ20.Re_reserved == Yes}
\end{split}
\end{equation}

\begin{equation} \label{equ:62}
\begin{split}
\cmdtt{E<> }& \cmdtt{deadlock &&}\\
& \cmdtt{S.Stream_transmission &&}\\
& \cmdtt{(BQ01.Re_reserved == No ||}\\
& \cmdtt{BQ11.Re_reserved == No) &&}\\
& \cmdtt{BQ20.Re_reserved == Yes}
\end{split}
\end{equation}

\begin{equation} \label{equ:63}
\begin{split}
& \cmdtt{S.Stream_transmission}\\
& \cmdtt{--> (BQ00.Re_reserved == Yes &&}\\
& \cmdtt{BQ01.Re_reserved != Yes &&}\\
& \cmdtt{BQ10.Re_reserved != Yes &&}\\
& \cmdtt{BQ11.Re_reserved != Yes &&}\\
& \cmdtt{BQ20.Re_reserved != Yes &&}\\
& \cmdtt{L0.Can_I_receive == Yes &&}\\
& \cmdtt{L1.Can_I_receive != Yes &&}\\
& \cmdtt{L2.Can_I_receive != Yes) ||}\\
& \cmdtt{(BQ00.Re_reserved != Yes &&}\\
& \cmdtt{BQ01.Re_reserved == Yes &&}\\
& \cmdtt{BQ10.Re_reserved == Yes &&}\\
& \cmdtt{BQ11.Re_reserved != Yes &&}\\
& \cmdtt{BQ20.Re_reserved != Yes &&}\\
& \cmdtt{L0.Can_I_receive != Yes &&}\\
& \cmdtt{L1.Can_I_receive == Yes &&}\\
& \cmdtt{L2.Can_I_receive != Yes) ||}\\
& \ \ \ \ \ \ \ \ \ \ \ \ \ \ \ \ \ \ \ \ \ \ \vdots\\
& \cmdtt{(BQ00.Re_reserved == Yes &&}\\
& \cmdtt{BQ01.Re_reserved == Yes &&}\\
& \cmdtt{BQ10.Re_reserved == Yes &&}\\
& \cmdtt{BQ11.Re_reserved == Yes &&}\\
& \cmdtt{BQ20.Re_reserved == Yes &&}\\
& \cmdtt{L0.Can_I_receive == Yes &&}\\
& \cmdtt{L1.Can_I_receive == Yes &&}\\
& \cmdtt{L2.Can_I_receive == Yes)}
\end{split}
\end{equation}

In the table included in the annex at the end of this document it is possible to see which queries are actually satisfied and which ones are not.

\section*{Acknowledgements}

This work is supported in part by the Spanish Agencia Estatal de Investigaci\'{o}n 
(AEI) and in part by FEDER funding through grant TEC2015-70313-R 
(AEI/FEDER, UE).

\onecolumn

\bibliographystyle{IEEEtran}
\bibliography{IEEEabrv,srp-csrp-models}

\newpage

\section*{Annex}
The following table summarises the results obtained when executing the presented queries in the corresponding model. It is important to note that the fact that a query is satisfied does not necessarily imply that termination and consistency are provided, just like the fact that a query is not satisfied does not imply the contrary. For more details on the queries please check Sections~\ref{TE}, \ref{CE}, \ref{CSRP_TE} and \ref{CSRP_CE}.

\begin{table}[!h]
\renewcommand{\arraystretch}{1.3}
\caption{Queries results.}
\label{QR}
\centering
\begin{tabular}{|c|c|c|c|c|c|}
 \hline
 Equation & Protocol & Result & Equation & Protocol & Result\\
 \hline
 \hline
 1 & SRP & \textcolor{LimeGreen}{Satisfied} & 
 33 & CSRP & \textcolor{LimeGreen}{Satisfied}\\
 2 & SRP & \textcolor{LimeGreen}{Satisfied} & 
 34 & CSRP & \textcolor{LimeGreen}{Satisfied}\\
 3 & SRP & \textcolor{LimeGreen}{Satisfied} & 
 35 & CSRP & \textcolor{LimeGreen}{Satisfied}\\
 4 & SRP & \textcolor{LimeGreen}{Satisfied} & 
 36 & CSRP & \textcolor{LimeGreen}{Satisfied}\\
 5 & SRP & \textcolor{LimeGreen}{Satisfied} & 
 37 & CSRP & \textcolor{LimeGreen}{Satisfied}\\
 6 & SRP & \textcolor{LimeGreen}{Satisfied} & 
 38 & CSRP & \textcolor{LimeGreen}{Satisfied}\\
 7 & SRP & \textcolor{LimeGreen}{Satisfied} & 
 39 & CSRP & \textcolor{LimeGreen}{Satisfied}\\
 8 & SRP & \textcolor{LimeGreen}{Satisfied} & 
 40 & CSRP & \textcolor{LimeGreen}{Satisfied}\\
 9 & SRP & \textcolor{LimeGreen}{Satisfied} & 
 41 & CSRP & \textcolor{LimeGreen}{Satisfied}\\
 10 & SRP & \textcolor{LimeGreen}{Satisfied} & 
 42 & CSRP & \textcolor{LimeGreen}{Satisfied}\\
 11 & SRP & \textcolor{LimeGreen}{Satisfied} & 
 43 & CSRP & \textcolor{LimeGreen}{Satisfied}\\
 12 & SRP & \textcolor{LimeGreen}{Satisfied} & 
 44 & CSRP & \textcolor{LimeGreen}{Satisfied}\\
 13 & SRP & \textcolor{LimeGreen}{Satisfied} & 
 45 & CSRP & \textcolor{LimeGreen}{Satisfied}\\
 14 & SRP & \textcolor{LimeGreen}{Satisfied} & 
 46 & CSRP & \textcolor{LimeGreen}{Satisfied}\\
 15 & SRP & \textcolor{LimeGreen}{Satisfied} & 
 47 & CSRP & \textcolor{LimeGreen}{Satisfied}\\
 16 & SRP & \textcolor{LimeGreen}{Satisfied} & 
 48 & CSRP & \textcolor{LimeGreen}{Satisfied}\\
 17 & SRP & \textcolor{LimeGreen}{Satisfied} & 
 49 & CSRP & \textcolor{LimeGreen}{Satisfied}\\
 18 & SRP & \textcolor{LimeGreen}{Satisfied} & 
 50 & CSRP & \textcolor{LimeGreen}{Satisfied}\\
 19 & SRP & \textcolor{LimeGreen}{Satisfied} & 
 51 & CSRP & \textcolor{red}{Not satisfied}\\
 20 & SRP & \textcolor{LimeGreen}{Satisfied} & 
 52 & CSRP & \textcolor{LimeGreen}{Satisfied}\\
 21 & SRP & \textcolor{LimeGreen}{Satisfied} & 
 53 & CSRP & \textcolor{LimeGreen}{Satisfied}\\
 22 & SRP & \textcolor{red}{Not satisfied} & 
 54 & CSRP & \textcolor{red}{Not satisfied}\\
 23 & SRP & \textcolor{LimeGreen}{Satisfied} & 
 55 & CSRP & \textcolor{LimeGreen}{Satisfied}\\
 24 & SRP & \textcolor{LimeGreen}{Satisfied} & 
 56 & CSRP & \textcolor{LimeGreen}{Satisfied}\\
 25 & SRP & \textcolor{red}{Not satisfied} & 
 57 & CSRP & \textcolor{red}{Not satisfied}\\
 26 & CSRP & \textcolor{LimeGreen}{Satisfied} & 
 58 & CSRP & \textcolor{LimeGreen}{Satisfied}\\
 27 & CSRP & \textcolor{LimeGreen}{Satisfied} & 
 59 & CSRP & \textcolor{LimeGreen}{Satisfied}\\
 28 & CSRP & \textcolor{LimeGreen}{Satisfied} & 
 60 & CSRP & \textcolor{LimeGreen}{Satisfied}\\
 29 & CSRP & \textcolor{LimeGreen}{Satisfied} & 
 61 & CSRP & \textcolor{red}{Not satisfied}\\
 30 & CSRP & \textcolor{LimeGreen}{Satisfied} & 
 62 & CSRP & \textcolor{red}{Not satisfied}\\
 31 & CSRP & \textcolor{LimeGreen}{Satisfied} & 
 63 & CSRP & \textcolor{LimeGreen}{Satisfied}\\
 32 & CSRP & \textcolor{LimeGreen}{Satisfied} & - & - & - \\
 \hline
\end{tabular}
\end{table}

\end{document}